\documentclass[sigconf, screen, noacm]{acmart}

\usepackage{kotex}
\usepackage{amsmath}
\usepackage{tabularx}

\usepackage{amssymb}
\usepackage{tikz}
\usepackage[shortlabels]{enumitem}
\usepackage{xspace}
\usepackage{macros} 

\newcommand{\redtext}[1]{\textcolor{red}{#1}}
\newcommand{\nelssa}[0]{\textsf{NELSSA}\xspace}
\newcommand{\niparagraph}[1]{\vspace{2pt}\noindent\textbf{#1}}
\newcommand*\circled[1]{\tikz[baseline=(char.base)]{
            \node[shape=circle,draw,text=white, font=\bfseries, fill=black, inner sep=0.6pt] (char) {#1};}}
\newcolumntype{Y}{>{\centering\arraybackslash}X}

\AtBeginDocument{%
  }

\setcopyright{none} 
\copyrightyear{2026}
\acmYear{}
\acmDOI{}
\acmConference[]{}

\setcopyright{none}
\renewcommand\footnotetextcopyrightpermission[1]{}
\begin{document}

\title{NELSSA: A GPU–PNM Heterogeneous System for Mixed-Length LLM Serving via Length–based Request Placement}

\author{Sookyung Choi}
\affiliation{%
\institution{SK hynix}
\country{Republic of Korea}
}
\email{sookyung.choi@sk.com}

\author{Seungyong Lee}
\affiliation{%
\institution{SK hynix}
\country{Republic of Korea}
}
\email{seungyong.lee@sk.com}

\author{Kangkyu Park}
\affiliation{%
\institution{SK hynix}
\country{Republic of Korea}
}
\email{kangkyu.park@sk.com}

\author{Yunseo Chun}
\affiliation{%
\institution{SK hynix}
\country{Republic of Korea}
}
\email{yunseo.chun@sk.com}

\author{Junseok Lee}
\affiliation{%
\institution{SK hynix}
\country{Republic of Korea}
}
\email{junseok3.lee@sk.com}

\author{Hyeongseok Gwak}
\affiliation{%
\institution{SK hynix}
\country{Republic of Korea}
}
\email{hyeongseok1.gwak@sk.com}

\author{Myunghyun Rhee}
\affiliation{%
\institution{SK hynix}
\country{Republic of Korea}
}
\email{myunghyun.rhee@sk.com}

\author{Euiseok Kim}
\affiliation{%
\institution{SK hynix}
\country{Republic of Korea}
}
\email{euiseok.kim@sk.com}

\author{Donguk Moon}
\affiliation{%
\institution{SK hynix}
\country{Republic of Korea}
}
\email{donguk.moon@sk.com}

\author{Kwangsik Shin}
\affiliation{%
\institution{SK hynix}
\country{Republic of Korea}
}
\email{kwangsik.shin@sk.com}

\author{Guseul Heo}
\affiliation{%
\institution{KAIST}
\country{Republic of Korea}
}
\email{gsheo@casys.kaist.ac.kr}

\author{Youngpyo Joo}
\affiliation{%
\institution{SK hynix}
\country{Republic of Korea}
}
\email{youngpyo.joo@sk.com}

\author{Hoshik Kim}
\affiliation{%
\institution{SK hynix}
\country{Republic of Korea}
}
\email{hoshik.kim@sk.com}

\author{Jongse Park}
\authornote{Corresponding author.}
\affiliation{%
\institution{KAIST}
\country{Republic of Korea}
}
\email{jspark@casys.kaist.ac.kr}

\begin{abstract}
Modern LLMs and their agentic applications are broadening the range of serving workloads, spanning context lengths from a few hundred tokens to hundreds of thousands.
As these requests frequently interleave within the same serving window, LLM serving systems must handle highly heterogeneous \emph{mixed-length} workloads.
Such mixed-length workloads expose fundamental inefficiencies in GPU-centric serving architectures, whose throughput depends on large, memory-constrained batches.
In this paper, we present \nelssa, an LLM serving system that integrates GPUs with real-world Processing-near-Memory (PNM) accelerator devices to efficiently support mixed-length workloads.
\nelssa employs length-based request placement to route short-context requests to GPUs and long-context requests to the PNM tier, incorporating runtime migration to accommodate dynamic context growth without recomputation.
We prototype \nelssa as an end-to-end system, implementing device-level sparse attention on PNM, GPU decode kernels, and a host-side runtime that orchestrates scheduling and cross-tier memory movement over a CXL-enabled infrastructure with RPC and RDMA support.
Across mixed-length LLM workloads, \nelssa improves decode throughput by up to 5.5$\times$ in tokens/sec and reduces P99 latency by up to 15$\times$ compared to GPU-only baselines.
Our end-to-end prototype and experimental results suggest that integrated GPU–PNM serving, enabled by CXL-based disaggregation, is a promising system paradigm for scalable and flexible LLM  infrastructures that support evolving workloads.
\end{abstract}

\keywords{LLM serving, Heterogeneous system, Processing-near-Memory (PNM), Mixed-length; Request placement}

\maketitle

\section{Introduction}
\label{sec:intro}

The recent surge of generative AI, initially driven by ChatGPT~\cite{chatgpt}, has evolved toward agentic AI workloads. 
AI coding assistants such as Claude Code~\cite{claude} and Codex~\cite{codex} exemplify this shift, demonstrating how LLMs increasingly function as autonomous agents capable of planning and executing complex tasks~\cite{agent1, agent2, agent3, agent4, SWEagent, SWEbench}.
As these agentic applications expand, LLM serving systems must support significantly more complex and longer-running workloads than in conventional chatbot deployments.

One prominent consequence of this shift is the emergence of \emph{mixed-length} LLM serving workloads. 
In production traces, request lengths span from a few hundred tokens to well beyond 100K tokens, while short and long requests frequently interleave within the same serving window~\cite{trace1, trace2, trace3, mooncake, medha, sarathi}. 
Such coexisting and unpredictable length variation forces modern LLM serving systems to manage  heterogeneous workloads under shared resources.
However, GPU-centric LLM serving architectures struggle with mixed-length workloads. 
Long-context requests consume a substantially large KV cache, quickly exhausting limited GPU memory capacity, which collapses the effective batch size and reduces throughput \cite{kvquant}.
When co-scheduled with short requests, long executions further induce head-of-line (HoL) blocking, inflating tail latency. 
Even a small fraction of long-context requests can therefore degrade overall system efficiency in GPU-only deployments.
To overcome these challenges, prior work has pursued two main directions. 
(1) \textbf{KV cache reduction techniques} shrink long-context memory footprints via compression, quantization, or pruning, but inevitably trade accuracy for capacity~\cite{streamingllm, H2O, kvzip, snapkv, pyramidkv, infiniplot, fastkv}. 
(2) \textbf{Memory disaggregation and offloading approaches} move KV caches beyond GPU's device memory to host memory or near-memory devices~\cite{flexgen, infinigen, retrievalattention, arkvale, retroinfer}, alleviating capacity pressure but introducing interconnect bandwidth bottlenecks. 

\begin{figure}[!t]
\centering
\includegraphics[width=0.95\columnwidth]{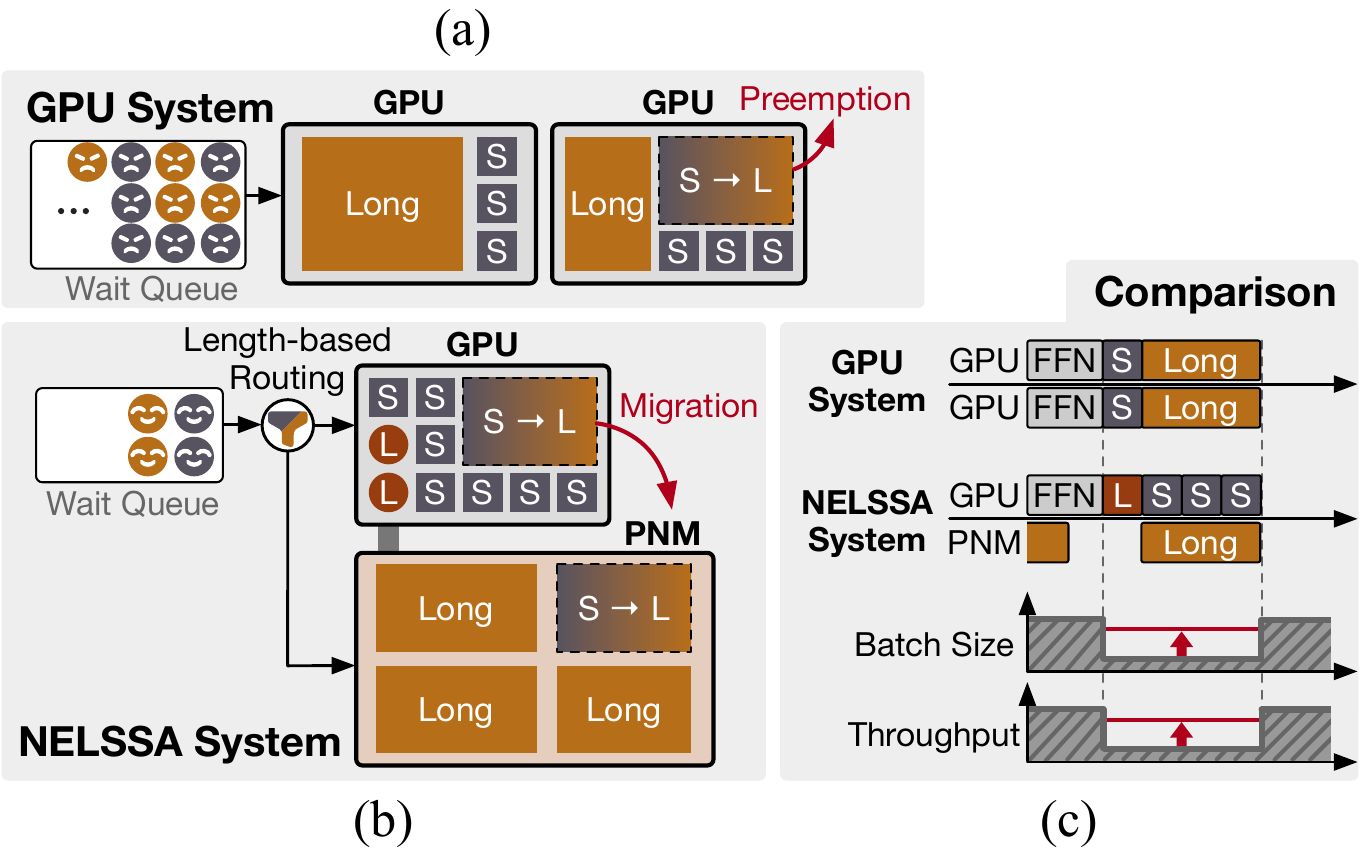}
\vspace{-2ex}
\caption{(a) GPU-only systems share mixed-length requests in a single batch, where long-context executions reduce batch size and induce head-of-line blocking.
(b) NELSSA separates requests via length-based placement, assigning short requests to GPUs and long requests to PNM, exploiting runtime migration to handle context growth.
(c) This heterogeneity maintains larger batch sizes and improves throughput by mitigating interference from long-context requests.}
\vspace{-5ex}
\label{fig:overview}
\end{figure}

Recently, a line of work has attempted to combine these two directions by leveraging sparse attention on Processing-near-Memory (PNM) accelerators~\cite{longsight, scalable_pnm}. 
By executing sparse attention near large-capacity memory, these designs promise to alleviate GPU memory capacity pressure, while reducing interconnect bandwidth overhead. 
However, these approaches exhibit fundamental limitations in supporting modern mixed-length LLM serving.
Consequently, sparse PNM-based designs face several key challenges:
\begin{itemize}[labelindent=0.3em,nolistsep,leftmargin=1.0em]
\item \textbf{Limitation 1: Long-only optimization.}
Existing sparse PNM solutions optimize long-context requests in isolation. In mixed-length scenarios with short and long requests interleaved, naively routing all requests to PNMs can degrade performance for short requests that are more efficiently executed on GPUs.
\item \textbf{Limitation 2: Lack of support for dynamic context growth.}
Most prior approaches assume static request lengths and determine offloading decisions upfront. However, modern agentic LLM workloads often exhibit unpredictable growth in generated context during decoding, rendering static routing insufficient.
\item \textbf{Limitation 3: Absence of full-stack system integration.}
Existing works typically focus on architectural feasibility or algorithmic acceleration without demonstrating an end-to-end system integrating real PNM hardware, host-side coordination, and device-level support required for practical LLM serving.
\end{itemize}
To tackle these limitations, we build \nelssa, an LLM serving system that integrates GPUs with real-world Processing-near-Memory (PNM) accelerator devices to efficiently support mixed-length workloads.
\nelssa employs length-based request placement with runtime migration to handle dynamic context growth, routing short requests to GPUs and long-context requests to PNMs through hardware–runtime co-design.
We prototype \nelssa as an end-to-end system, implementing device-level sparse attention on PNM, GPU decode kernels, and a host-side runtime that orchestrates scheduling and cross-tier memory movement over a CXL-enabled infrastructure with RPC and RDMA support.
Figure~\ref{fig:overview} illustrates the differences between GPU-only systems and \nelssa in handling mixed-length workloads.
Our contributions are as follows:
\begin{description}[leftmargin=1.5em]
\item \textbf{(1) Length-based request placement with hardware-aware routing.}
We design a routing mechanism that assigns short requests to GPUs and long-context requests to a sparse PNM tier based on a hardware-derived crossover threshold. 
Unlike heuristic or static partitioning, our approach analytically determines the placement boundary from device-level performance characteristics, enabling each request to execute on the most suitable compute tier. 
This split-batch hybrid routing maximizes effective GPU batch size, while offloading memory-capacity-bound long requests to PNM, improving throughput and mitigating head-of-line blocking under mixed-length workloads.
\item \textbf{(2) Runtime migration for dynamic context growth.}
We introduce a lightweight migration mechanism that enables requests to transition from GPU to PNM execution as their context length grows during decoding. 
Rather than relying on recomputation or swap-based eviction when memory pressure arises, \nelssa performs a one-way handoff that preserves KV cache and avoids service interruption.
This design ensures stable throughput and latency even under unpredictable, dynamically expanding workloads such as reasoning-heavy or agent-driven inference.
\item \textbf{(3) End-to-end GPU–PNM heterogeneous system integration.}
We build an end-to-end GPU–PNM heterogeneous LLM serving system integrating CXL-attached PNM devices, device-level sparse attention engines, and a host-side control plane. 
Our implementation spans the full execution path, including RDMA communication, gRPC/TCP control, KV cache management across GPU and multi-PNM nodes, and distributed attention aggregation over a CXL interconnect. 
Deployed on real hardware, \nelssa demonstrates tightly coupled GPU–PNM co-execution, establishing heterogeneous memory–compute integration as a viable architecture for large-scale LLM serving, while enabling flexible and scalable disaggregated deployment through CXL.
\end{description}

To evaluate \nelssa, we conduct end-to-end experiments using realistic mixed-length serving traces that capture highly skewed and interleaved short and long-context requests. 
Compared to a state-of-the-art GPU-only serving system, \nelssa improves decode throughput by up to 5.5$\times$ in tokens/sec and reduces P99 latency by 15$\times$, while maintaining the same level of accuracy as the GPU-only baseline. 
These gains stem from restoring effective running batch size on GPUs through length-based heterogeneous placement and mitigating head-of-line blocking caused by long-context requests.
Even under dynamically growing contexts, \nelssa sustains stable throughput without recomputation overhead.
As emerging LLM workloads increasingly exhibit diverse and unpredictable context lengths, GPU-only infrastructures remain rigid and ill-suited to support such heterogeneity. 
By combining hardware heterogeneity, runtime scheduling, and CXL-based disaggregation, \nelssa presents a system design that enables scalable and flexible resource composition for mixed-length LLM serving. 
Our end-to-end prototype and evaluation results suggest this design as a practical approach to serving evolving LLM workloads.
\section{Motivation}\label{sec2}

\subsection{The Mixed-Length Problem in LLM Serving}\label{sec2.1}
As LLMs with support for increasingly long contexts have become widely deployed, real-world serving workloads have come to exhibit a mix of short and long requests. Such mixed-length distributions have been observed in prior workload analyses \cite{sarathi, medha}; for instance, the Mooncake conversation trace \cite{mooncake} shows that approximately 34\% of requests fall under 4K tokens while requests exceeding 100K tokens are served concurrently. As LLM usage expands into more complex workloads—such as multi-agent systems, chain-of-thought reasoning \cite{cot}, and agentic coding--the proportion and length of long-context requests are expected to grow further.

\begin{figure}[!t]
\centering
\includegraphics[width=1.0\columnwidth]{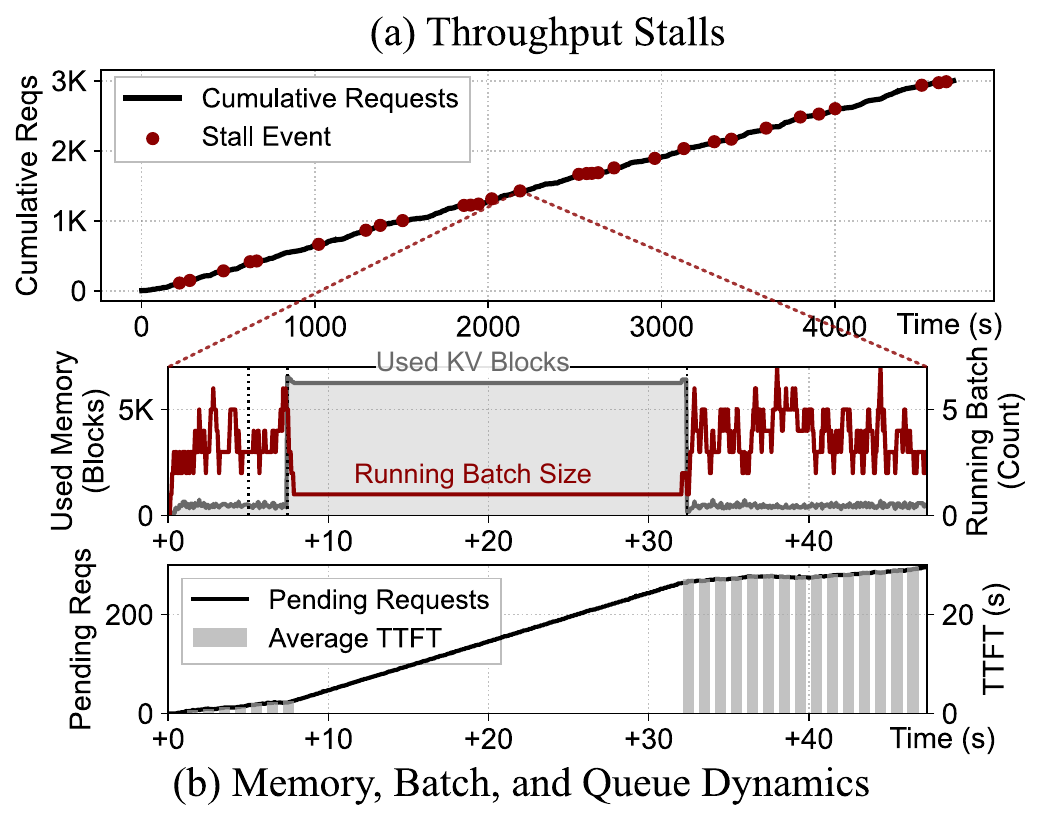}
\caption{GPU memory saturation, batch size collapse, and throughput stalls under mixed-length workload on a single H100 GPU (94 GB).}
\label{fig_hol}
\end{figure}

When such heterogeneous requests share the same serving environment, inefficiency arises. A long-context request occupies a substantial portion of GPU HBM with its KV cache, preventing the scheduler from admitting new requests into the batch and, in memory-saturated conditions, forcing the running batch size to 1. This triggers Head-of-Line (HoL) blocking: subsequent short requests stall in the queue until the long request completes, leading to tail latency spikes and pending queue buildup. 
Figure~\ref{fig_hol} illustrates this effect on a single H100 GPU (94 GB), where scheduling  a long-context request under HBM capacity pressure causes memory saturation, batch size collapse, and periodic throughput stalls.

The root cause lies in a fundamental resource mismatch between long-context decode attention and GPU HBM-based infrastructure. The decoding phase of Transformer-based LLM inference is auto-regressive, requiring the model to access the full accumulated KV cache to generate each token—rendering it a heavily memory-bound workload with a high memory-to-compute ratio \cite{david, roofline, flashattention, hpu, scaling_transformer}. As context length grows, this effect intensifies: long-context decode attention enters a capacity-bound regime in which the KV cache simultaneously saturates both HBM capacity and bandwidth, while GPU compute utilization decreases \cite{hetero}. 
Scaling out GPUs does not resolve this efficiently \cite{gpuscaling1, gpuscaling2, gpuscaling3}, as adding GPUs increases total HBM capacity but scales compute and memory together, resulting in proportionally higher cost and power draw while leaving an increasing fraction of compute underutilized.

These observations point to a need for a dedicated memory tier that can provide sufficient capacity and bandwidth for long-context decode attention in a cost-effective manner, and that integrates seamlessly with existing GPU infrastructure. However, deploying such a tier in practice introduces two additional challenges that arise in real serving environments: the coexistence of heterogeneous short and long requests, and the unpredictable growth of context length during decoding. We analyze these in the following sections.


\subsection{Limitations of Existing Approaches}\label{sec2.2}
As analyzed in Section~\ref{sec2.1}, GPU-only infrastructure introduces structural inefficiencies in long-context decode attention. Prior work has pursued two main directions to address this: reducing the KV cache memory footprint through algorithmic techniques, and offloading KV caches to memory outside the GPU. However, neither direction sufficiently addresses the challenges that arise in mixed-length serving environments.

\niparagraph{KV cache reduction.} KV compaction and quantization techniques \cite{streamingllm, snapkv, pyramidkv, duoattention, kvzip, infiniplot, fastkv} aim to reduce KV cache size to fit within GPU HBM capacity. However, since the tokens relevant to a given decode query cannot be known in advance, these methods discard information in a query-agnostic manner, leading to accuracy degradation. Aggressive compression amplifies this risk, creating a difficult tradeoff between capacity headroom and accuracy loss.

\niparagraph{Sparse attention on GPU.}
Applying sparse attention directly on GPUs can reduce attention computation and memory bandwidth. However, accuracy-preserving sparse attention still requires the full KV cache to remain resident in HBM because token selection is query-dependent. Therefore, GPU-side sparsity alone does not remove the capacity pressure that collapses batch size and induces head-of-line blocking under mixed-length workloads.


\niparagraph{CPU offloading with sparse attention.} 
Approaches such as InfiniGen \cite{infinigen} and RetroInfer \cite{retroinfer} retain the full KV cache in CPU DRAM and transfer a sparsely selected subset to the GPU via PCIe for each decode step. By preserving the full KV cache, these methods avoid the accuracy loss inherent in compaction. However, PCIe bandwidth poses a fundamental bottleneck regardless of the selection ratio chosen. Increasing the selection ratio improves accuracy but raises bandwidth pressure, while decreasing it reduces data movement but degrades accuracy. Furthermore, since attention computation ultimately executes on the GPU, the per-step transfer of selected KV vectors across PCIe cannot be avoided. Even under aggressive sparsity, this KV-vector-scale data movement sustains non-trivial bandwidth pressure on every decode step.

\niparagraph{Sparse attention on PNM.} PNM architectures have been explored as a promising direction to overcome these limitations. Early efforts such as CXL-PNM \cite{cxl_pnm} and Hermes \cite{hermes} demonstrated strong energy efficiency by executing dense attention directly on PNM. However, simply relocating dense attention execution does not escape the capacity-bandwidth dilemma. The large-capacity memory required for long-context serving inherently offers lower bandwidth than GPU HBM, imposing a fundamental constraint on decode throughput. To bridge this gap, subsequent work proposed combining sparse attention with PNM execution \cite{cal_nelssa, longsight, scalable_pnm}, demonstrating that sparsity can effectively offset the bandwidth disadvantage and enabling large-scale long-context serving.

\begin{figure}[!t]
\centering
\includegraphics[width=1.0\columnwidth]{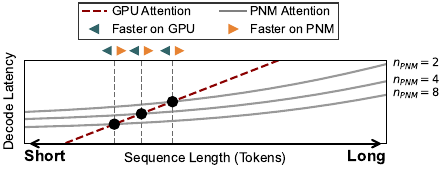}
\caption{Decode attention latency across sequence lengths. A crossover point exists beyond which PNM execution can outperform GPU execution, with its location varying across hardware configuration and system conditions.}
\label{fig_cross}
\end{figure}

However, existing PNM-based approaches uniformly route all requests to PNM regardless of length, which is not always the most efficient strategy. As shown in Figure~\ref{fig_cross}, a crossover point exists between GPU dense attention and PNM sparse attention as a function of sequence length. Below this threshold, the fixed overhead of PNM offloading, including network latency and host-side dispatch, outweighs the bandwidth advantage PNM provides, making the GPU the faster execution path. Naively routing all requests to PNM can therefore introduce performance regression for short requests, which constitute the majority of real-world workloads.

In short, each approach offers advantages under specific conditions but falls short as a complete solution for mixed-length serving. GPU-only execution is prone to HoL blocking from long requests; algorithmic approaches incur either accuracy loss or interconnect bottlenecks; and PNM-only routing can degrade performance for short requests. Taken together, these observations point to the need for a length-based placement strategy that adaptively assigns each request to its most suitable execution tier, routing short requests to the GPU and long requests to PNM.

\subsection{The Challenge of Dynamic Context Growth}\label{sec2.3}
The approaches in Section~\ref{sec2.2} implicitly assume that the context length of a request is known before execution begins. While this holds in many scenarios, it does not always reflect the reality of modern serving workloads. Workloads such as chain-of-thought reasoning, multi-step agentic tasks, and complex inference often begin with relatively short inputs whose context grows substantially during decoding \cite{reasoning1, reasoning2, reasoning3, reasoning4, reasoning5}, in ways that are difficult to anticipate in advance \cite{response, output1, output2, output3, output4, output5, output6, output7}. When this growth causes GPU memory to be exhausted mid-execution, the scheduler must preempt the request.

Upon preemption, two recovery strategies are available. Recomputation discards the evicted KV cache and regenerates it from scratch when the request is rescheduled, consuming GPU computing resources on recovery rather than on new token generation. Swap retains the evicted KV cache in CPU DRAM but requires transferring it back over PCIe before inference can resume, reintroducing the very interconnect pressure that offloading sought to avoid \cite{instinfer}. Neither strategy resolves the underlying problem without significant overhead. Notably, vLLM \cite{vllm} defaults to recomputation over swapping, as the system-level overhead of PCIe-based KV transfer outweighs its recovery latency benefit.

The fundamental limitation shared by both strategies is that execution remains anchored on the GPU. Regardless of where the KV cache is stored, it must be returned to the GPU before inference can resume. This points to the need for an approach that goes beyond storage relocation, one in which execution itself transitions away from the GPU so that attention can be performed directly where the KV cache resides, without a return trip. We develop this direction in Section~\ref{sec3}.

\section{Design Principles and Architecture Overview}\label{sec3}
The analyses in Section~\ref{sec2} converge on a single conclusion. Long-context decode attention is inherently memory-bound, making it structurally ill-suited to GPU HBM-centric infrastructure. It saturates HBM capacity while leaving tensor core utilization low, exposing an efficiency mismatch that compounds as context length grows. Critically, this is not a problem of \textit{where to store} the KV cache. It is a problem of \textit{where to execute} attention, and fundamentally, an *execution placement* problem. \nelssa begins from this perspective, treating PNM as a dedicated execution tier for long-context attention rather than an auxiliary offload target. Attention computation completes entirely within PNM and does not return to the GPU.

\begin{figure}[!t]
\centering
\includegraphics[width=1.0\columnwidth]{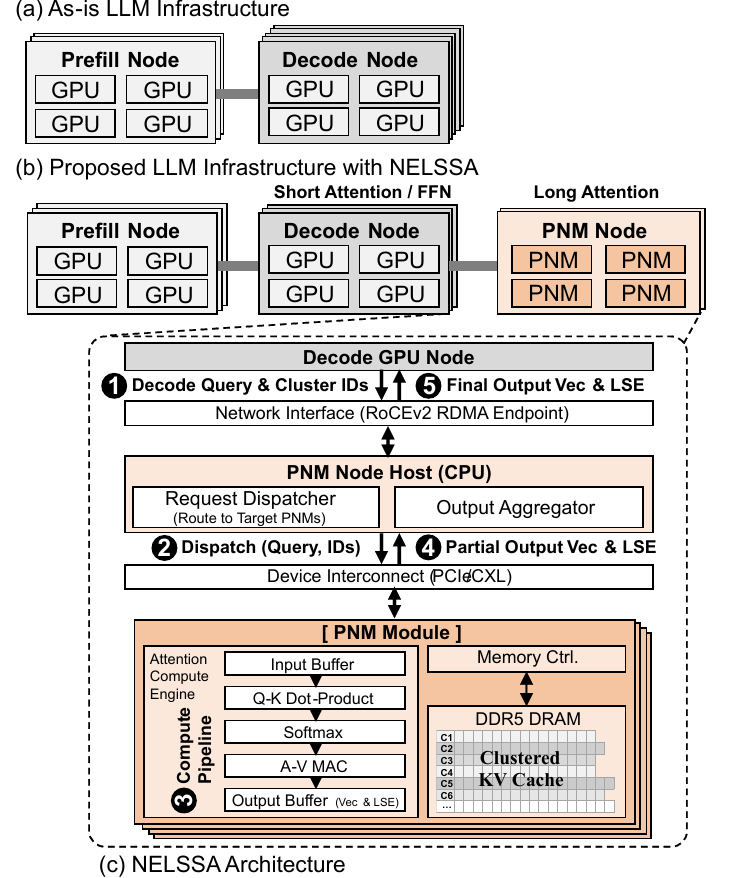}
\caption{Overall architecture of \nelssa.}
\vspace{-5pt}
\label{fig_infra}
\end{figure}

This assignment is effective for two reinforcing reasons. First, long-context decode attention is inherently memory-bound, which aligns well with PNM architecture. Each decode step requires reading the full accumulated KV cache with low arithmetic intensity, matching the characteristics of large-capacity memory with data-proximate access. Second, sparse attention structurally complements PNM's lower memory bandwidth relative to GPU HBM, offsetting the bandwidth gap and making the combination more capable than either alone. Based on this, \nelssa assigns each tier to the workload it handles best. The GPU executes dense attention for short requests and FFN computations for all requests, while PNM exclusively handles sparse attention for long-context requests. This specialization is governed by two design principles. First, execution placement must be determined by request length. GPU-only execution is susceptible to HoL blocking from long requests, while routing all requests to PNM degrades performance for short ones, and no single strategy handles the full spectrum efficiently. Second, once execution transfers to the PNM tier, it must not return to the GPU. Prior approaches differ in how much data they move, but share the structural constraint that execution ultimately resumes on the GPU. \nelssa eliminates this constraint entirely.

\nelssa realizes these two principles through three mechanisms:
\begin{itemize}[labelindent=0.3em,nolistsep,leftmargin=1.0em]
\item \textbf{Length-aware hybrid routing (Section~\ref{sec4.1}-\ref{sec4.2})}: routes short requests to GPU FlashAttention and long requests to PNM sparse attention, with the placement boundary derived from hardware performance characteristics.
\item \textbf{Seamless background migration (Section~\ref{sec4.3})}: transfers the KV cache of dynamically growing requests to PNM via a background stream, without interrupting ongoing token generation.
\item \textbf{GPU–PNM heterogeneous system architecture (Section~\ref{sec5})}: realizes both mechanisms on real PNM hardware within an end-to-end integrated serving stack.
\end{itemize}

The overall system architecture is illustrated in Figure~\ref{fig_infra}.

\section{Length-Aware Execution Placement}\label{sec4}
\subsection{Routing Threshold Derivation}\label{sec4.1}
The first design principle established in Section~\ref{sec3} calls for placing decode requests on different execution tiers based on sequence length, which requires a concrete placement boundary. This boundary, the routing threshold $T_{\text{input}}$, determines which requests are handled on the GPU and which are offloaded to the PNM tier. Rather than fixing this boundary heuristically, \nelssa derives $T_{\text{input}}$ from the hardware performance characteristics of each execution path, providing a principled baseline that captures when PNM execution offers a more suitable operating point than GPU execution across latency, throughput, and efficiency considerations. To this end, \nelssa constructs a decode attention latency model for both paths and identifies their crossover point as $T_{\text{input}}$.

\niparagraph{GPU dense attention.} During the decode phase, each generated token requires reading the full accumulated KV cache, making attention strictly memory-bound \cite{roofline}. The latency of GPU attention is dominated by HBM bandwidth and can be approximated as:

\begin{equation}\label{eqn:eq1}
T_{\text{GPU}}(L) = \frac{L \cdot M_{kv}}{BW_{\text{GPU}}}
\end{equation}

where $L$ is the sequence length, $M_{kv}$ the KV cache size per token, and $BW_{GPU}$ the HBM bandwidth of the GPU.

\niparagraph{PNM sparse attention.} For \nelssa with $N$ PNM devices, the decode latency is approximated as the sum of three dominant components: GPU-side centroid similarity search, distributed sparse attention across $N$ PNM devices, and communication overhead $T_{comm}(N)$:

\begin{equation}\label{eqn:eq2}
T_{\text{PNM}}(L, N) = \frac{L \cdot M_{kv}}{C \cdot BW_{\text{GPU}}} + \frac{L \cdot S \cdot M_{kv}}{N \cdot BW_{\text{PNM}}^{\text{eff}}} + T_{\text{comm}}(N)
\end{equation}

where $C$ is the compression factor of the centroid search, $S$ the sparse selection ratio, and $BW_{\text{PNM}}^{\text{eff}}$ the effective per-device bandwidth measured on our prototype under sparse random access. The communication term $T_{\text{comm}}(N)$ captures three additive costs: a fixed base network round-trip latency that applies to all configurations, a per-device incremental transfer cost that scales modestly with $N$, and a constant host-side dispatch and aggregation overhead that arises only in multi-PNM configurations.

\niparagraph{Crossover point.} Equating $T_{GPU}(L)$ and $T_{PNM}(L, N)$ yields a hardware-derived crossover point as a function of device count:

\begin{equation}
L_{\text{crossover}}(N) = 
\frac{T_{\text{comm}}(N)}{%
M_{kv} \left( 
\dfrac{C-1}{C \cdot BW_{\text{GPU}}} - 
\dfrac{S}{N \cdot BW_{\text{PNM}}^{\text{eff}}} 
\right)}
\end{equation}

This formulation shows that the bandwidth gap between HBM and DDR5 can be offset by sparsity and multi-PNM parallelism as long as $\frac{S}{N \cdot BW_{\text{PNM}}^{\text{eff}}} < \frac{C-1}{C \cdot BW_{\text{GPU}}}$, a condition that becomes progressively easier to satisfy as $N$ increases. 
Figure~\ref{fig_cross} illustrates how $L_{\text{crossover}}(N)$ varies with device count.
With a single PNM device, the crossover lies at a sequence length that exceeds most practical workloads, motivating the multi-PNM design. As $N$ increases, the crossover shifts toward shorter sequences, enabling a broader range of requests to be efficiently served on the PNM tier.

The purpose of this model is not to predict exact latency, but to expose the hardware-dependent trade-offs that govern when execution should transition across tiers. 
Applying the model to the prototype determines a hardware-specific crossover point, which \nelssa adopts as the routing threshold $T_{\text{input}}$. The exact value varies with hardware characteristics and deployment configuration.

\begin{figure}[!t]
\centering
\includegraphics[width=1.0\columnwidth]{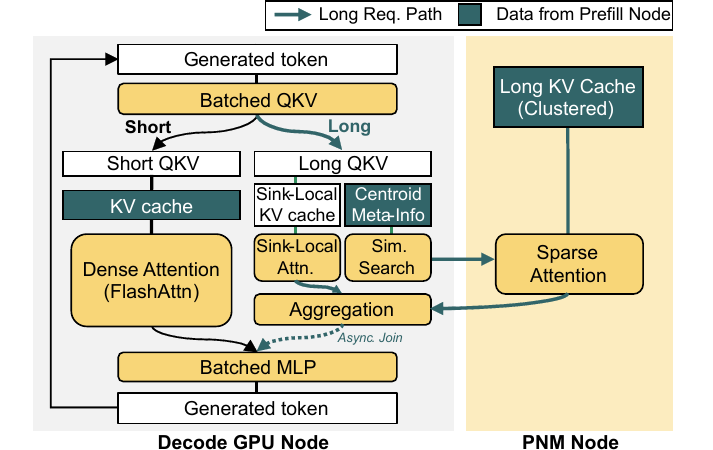}
\caption{Split-batch hybrid routing branches only the attention phase by sequence length, while QKV and MLP computations are globally batched on the GPU.}
\label{fig_routing}
\end{figure}

\subsection{Split-Batch Hybrid Routing}\label{sec4.2}
Based on the threshold $T_{input}$ derived in Section~\ref{sec4.1}, \nelssa partitions incoming requests into two execution paths. Short requests—those with sequence length below $T_{input}$—are executed on the GPU via FlashAttention, while long requests exceeding $T_{input}$ are routed to the PNM tier for sparse attention. This separation is enforced as an architectural invariant, not merely a routing policy: once a long request's attention is assigned to the PNM tier, it executes exclusively there without GPU recall. This structurally eliminates interconnect round-trips that would otherwise reintroduce interference to concurrently running short requests.

\niparagraph{Batched execution.} Separating the two execution paths must not come at the cost of GPU compute efficiency. To preserve batch utilization, \nelssa globally batches the QKV and MLP computations of both short and long requests on the GPU; the two paths diverge only at the attention stage. Short requests execute FlashAttention entirely within the GPU. For long requests, computation bifurcates: the GPU handles a local attention window covering sink and recent tokens \cite{flashattention}, while cluster-based sparse attention over the global context is asynchronously offloaded to the PNM. The sparse attention leverages query-driven cluster selection \cite{retroinfer}, and the partial results from GPU local attention and PNM global attention are merged via online softmax-based distributed aggregation—a combination that is mathematically equivalent to attention over the union of the local and selected sparse contexts via LSE-based merging. While PNM offloading proceeds, the GPU immediately continues with MLP computations for short requests; once the PNM result returns, the long request rejoins the next available MLP batch. This ensures that the execution of short requests is never stalled by PNM offloading of long requests.

The overall structure is illustrated in Figure~\ref{fig_routing}. This design maximizes tensor core utilization while providing the memory capacity required for long-context execution

\begin{figure}[!t]
\centering
\includegraphics[width=1.0\columnwidth]{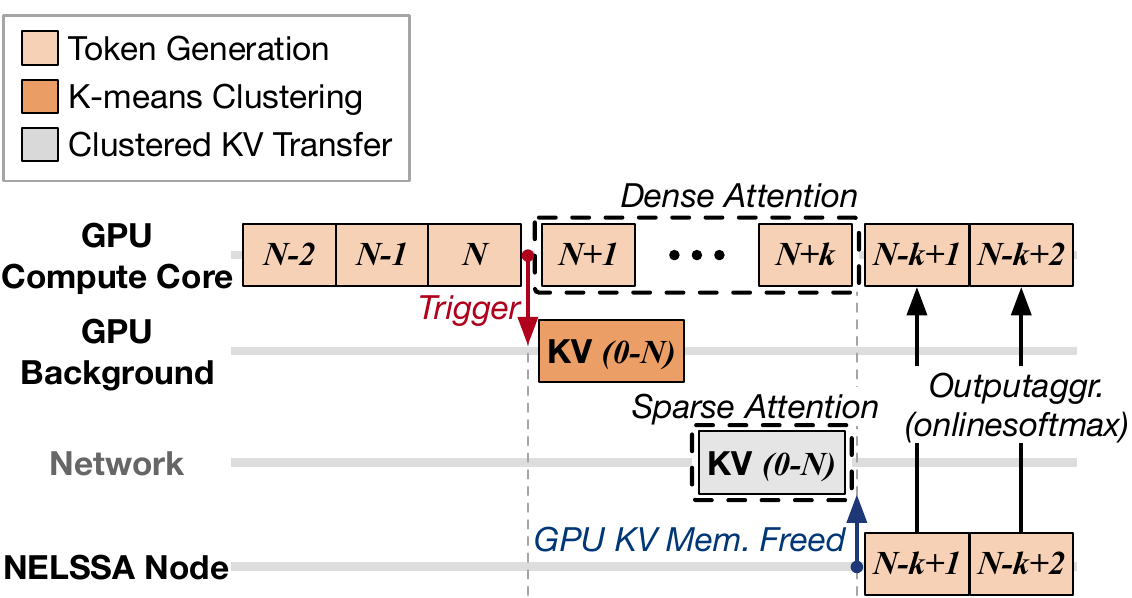}
\caption{Seamless background migration. Historical KV caches are asynchronously transferred to the PNM Node while GPU token generation continues uninterrupted.}
\vspace{-4mm}
\label{fig_mig}
\end{figure}

\subsection{Seamless Background Migration}\label{sec4.3}
The split-batch routing described in Section~\ref{sec4.2} determines the execution tier based on the input length of each request. However, as analyzed in Section~\ref{sec2.3}, context length is often not fixed at the time a request begins execution. A request that starts below $T_{input}$ and is initially assigned to the GPU can accumulate a growing KV cache during decoding, dynamically giving rise to the same HBM capacity pressure identified in Section~\ref{sec4.1}. To handle this, \nelssa extends the one-way execution handoff principle to dynamically growing requests, transferring the accumulated KV cache to the PNM tier in the background and executing sparse attention there without returning to the GPU.

\niparagraph{Migration logic.} \nelssa employs a threshold-based migration policy triggered by GPU memory pressure. A request becomes a migration candidate once its accumulated context length exceeds $T_{input}$.
Among all candidates, migration is initiated for the request with the largest accumulated context when preemption is imminent, ensuring that memory is reclaimed without disrupting requests that can continue executing on the GPU. This design avoids premature migration under sufficient memory headroom while preventing preemption-induced recomputation overhead.

\niparagraph{Background handoff.} Upon triggering, the GPU asynchronously transfers the historical KV cache to the PNM tier via a background stream, without interrupting token generation. Once the transfer completes, the request transitions to \nelssa's distributed execution mode. At handoff, the GPU retains only the lightweight Centroid Meta-Info required for subsequent routing and immediately reclaims the historical KV memory to accommodate new requests. KV cache generated by tokens produced after the handoff is periodically clustered in predefined chunks and appended to the PNM tier incrementally, amortizing clustering overhead and minimizing memory fragmentation. By consistently applying the one-way handoff principle to both statically classified long requests and dynamically growing ones, \nelssa separates long-context attention from the GPU in both cases, preserving overall system memory efficiency and throughput. The flow is shown in Figure~\ref{fig_mig}.

\section{GPU–PNM Heterogeneous System}\label{sec5}
\subsection{\nelssa System Architecture}\label{sec5.1}
The length-based routing and background migration mechanisms in Section~\ref{sec4} presuppose tight GPU–PNM interaction, and this section presents the architecture that makes them practically realizable.
This section presents the system architecture that makes these mechanisms practically realizable. Building on prefill-decode disaggregation \cite{splitwise, distserve}—a direction widely adopted in recent LLM serving systems—\nelssa introduces a disaggregated hybrid architecture that further separates long-context workloads onto a dedicated high-capacity memory tier, addressing the decode attention capacity bottleneck that prefill-decode separation alone cannot resolve (Figure~\ref{fig_infra}(b)).

The system comprises three components. The Prefill Node handles prefill computation for all requests. For long-context requests, it performs segmented K-means clustering in parallel with prefill computation, preparing the KV cache in a clustered form suitable for PNM offloading. This clustering scheme is adopted from RetroInfer \cite{retroinfer}, which reports that the associated overhead accounts for under approximately 2\% of total prefill time. Short-context requests are forwarded directly to the Decode GPU Node upon prefill completion. The Decode GPU Node executes FlashAttention for short-context requests and all FFN computations. For long-context requests, it retains only the lightweight Centroid Meta-Info and performs similarity search to derive the target cluster IDs for routing to the \nelssa Node. The \nelssa Node is a high-capacity memory node connected via RoCEv2, housing multiple PNM modules. Each module stores its assigned partition of the clustered KV cache in DDR5 DRAM following a head-wise partitioning scheme (Section~\ref{sec5.3}), and executes sparse attention in parallel. 

\niparagraph{Architectural isolation.} In this disaggregated infrastructure, the PNM Node must concurrently ingest massive prefill datasets and execute decode attention streams. Rather than relying on a complex runtime scheduler to dynamically balance these workloads, \nelssa enforces strict architectural isolation. Prefill data traffic is managed through an independent host-side memory path, bypassing the PNM's internal compute pipelines. This structural design prevents data-ingestion traffic from introducing latency jitter into active decode attention execution.

\niparagraph{Independent scaling.}
Beyond functional separation, the disaggregated structure enables independent scaling of each tier. Since the Decode GPU Node and the \nelssa Node connect over commodity Ethernet (RoCEv2), each tier scales independently. As the proportion of long-context workloads grows, additional \nelssa Nodes expand both memory capacity and attention throughput; when GPU-side compute becomes the bottleneck, Decode GPU Nodes scale out instead. 
In this way, disaggregation serves as a structural enabler that allows each tier to be provisioned according to its own workload characteristics.
\vspace{-1mm}

\subsection{Execution Pipeline}\label{sec5.2}
\nelssa's execution pipeline consists of an initial Prefill Phase and an iterative Decode Phase. The routing decision that determines whether a request is handled on the GPU or offloaded to the \nelssa Node is derived in Section~\ref{sec4.1}–\ref{sec4.2}; here we describe the execution flow for each request type once that assignment has been made.

\niparagraph{Prefill phase}: For long-context requests, segmented K-means clustering is performed layer-by-layer, overlapping with prefill computation to amortize the clustering overhead. The resulting Clustered KV Cache is offloaded to the \nelssa Node, while the GPU retains only the lightweight Centroid Meta-Info for subsequent routing.

\niparagraph{Decode phase}: During token generation, the GPU performs a similarity search using the Centroid Meta-Info to derive the Top-K target cluster IDs, transmitting them alongside the query to the \nelssa Node (\circled{1}). \nelssa's Host CPU Dispatcher distributes computation commands to the specific PNM modules holding the target data (\circled{2}). Each PNM module retrieves the target clusters from DDR5 DRAM into its input buffer and executes sparse attention locally, producing partial output vectors and Log-Sum-Exp (LSE) values (\circled{3}), \circled{4}). The Host's Output Aggregator collects these results and returns them to the GPU (\circled{5}), which then merges them with its own local attention results via online softmax-based LSE merging to generate the final token.

\niparagraph{Dynamic context update}: Whenever newly generated tokens accumulate to a predefined chunk size in GPU memory, the GPU independently clusters this chunk, appends it to the \nelssa Node, and updates the Centroid Meta-Info. This incremental approach amortizes clustering overhead and accommodates dynamic context growth while minimizing memory fragmentation.


\begin{figure}[!t]
\centering
\includegraphics[width=1.0\columnwidth]{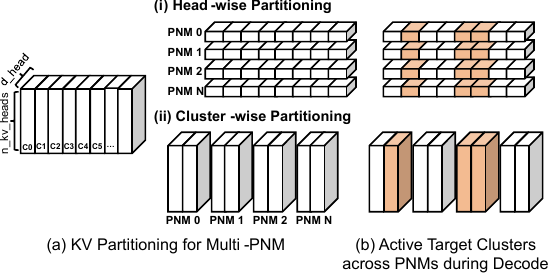}
\caption{KV cache partitioning strategies for multi-PNM development: (i) head-wise and (ii) cluster-wise.}
\label{fig_partitioning}
\end{figure}

\subsection{KV Partitioning for Multi-PNM Scalability}\label{sec5.3}
\nelssa distributes the KV cache across multiple PNM modules to scale both memory capacity and attention throughput in proportion to deployment needs. The partitioning strategy is a key design choice that determines the efficiency of the multi-PNM system (see Figure~\ref{fig_partitioning}). 
Cluster-wise partitioning (Figure~\ref{fig_partitioning}(ii)) stores complete clusters on individual modules, preserving large access granularity. However, depending on workload characteristics and sparse selection patterns, target clusters may concentrate on specific modules, potentially leading to load imbalance. 
While this effect may diminish as context length grows, \nelssa adopts head-wise partitioning (Figure~\ref{fig_partitioning}(i)), where all PNM modules process the same list of clusters but compute only their assigned attention heads. This provides robust load balancing regardless of context length or dynamic sparsity patterns.
Although partitioning across multiple modules reduces the access granularity per cluster, this size still exceeds the typical row buffer capacity (1–2 KB) of DDR5 DRAM, sufficiently amortizing the row activation penalty (t\_{RCD}). Combined with bank-level parallelism, head-wise partitioning maintains high effective bandwidth even under random access patterns.



\section{Implementation}\label{sec6}
The architecture in Section~\ref{sec4} is a reference design targeting a dedicated compute pipeline. Our prototype realizes it on real PNM hardware with DDR5 DRAM and ARM Neoverse V2 cores over CXL. 
Using these general-purpose cores in place of the dedicated compute units, the prototype provides a conservative characterization of end-to-end behavior and enables direct measurement of the system-level properties central to this work.

\begin{figure}[!t]
\centering
\includegraphics[width=1.0\columnwidth]{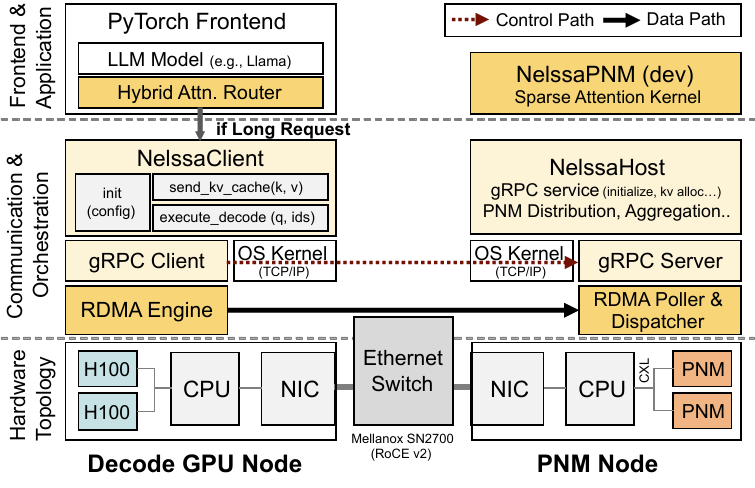}
\caption{Overall \nelssa SW stack.}
\label{fig_fullstack}
\end{figure}

\begin{figure}[!t]
\centering
\includegraphics[width=1.0\columnwidth]{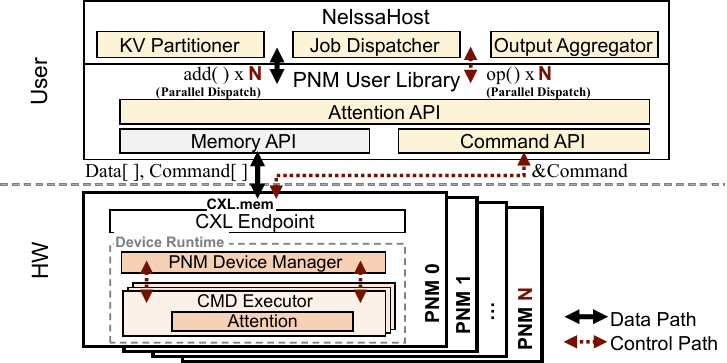}
\caption{\nelssa PNM software stack.}
\label{fig_pnm_dev}
\end{figure}


\begin{figure*}[!t]
\centering
\includegraphics[width=1.0\linewidth]{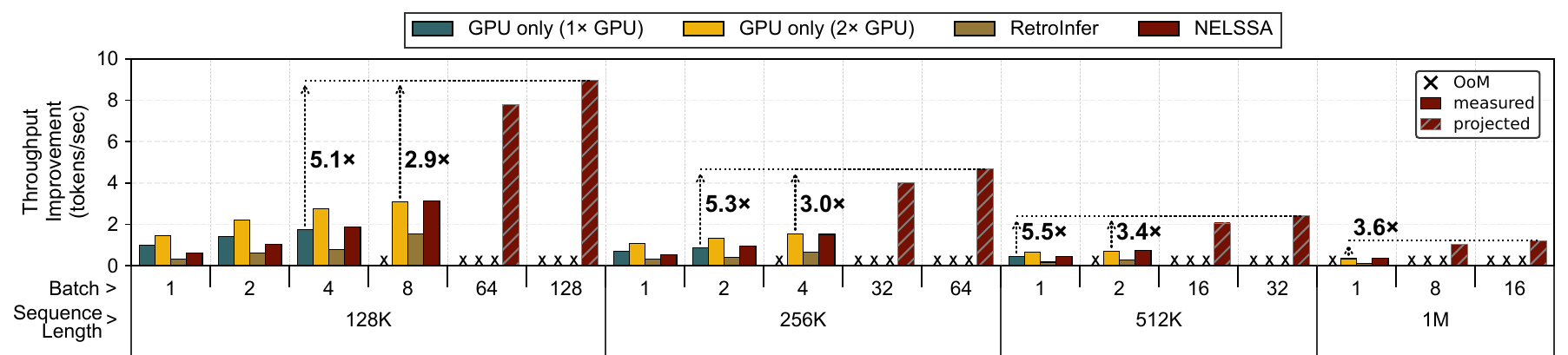}
\caption{Decode throughput comparison across varying sequence lengths and batch sizes (measured/projected). \nelssa achieves up to 5.5$\times$ higher throughput, including configurations where GPU-only baselines fail due to out-of-memory (OoM).}
\label{fig_throughput3}
\end{figure*}

\subsection{End-to-End System Integration}
We implemented an end-to-end prototype system connecting the Decode GPU Node and the PNM Node using real hardware and open-source frameworks (see Figure~\ref{fig_fullstack}). The frontend on the GPU Node is implemented in Python for integration with PyTorch, while the performance-critical communication backend is implemented in C++. 
To avoid the latency and CPU overhead of routing all traffic through a kernel-managed network stack, \nelssa decouples the control path from the data path. Latency-insensitive operations such as session initialization and KV cache allocation are handled via gRPC over TCP/IP, while performance-critical transfers, specifically KV cache offloading and query/output vector exchanges, are executed over RoCEv2 RDMA, bypassing the OS kernel and enabling direct memory-to-memory transfers. A dedicated RDMA Poller thread eliminates context-switching delays, achieving a one-way latency of a few microseconds. 
On the PNM Node, the NelssaHost dispatches commands to the target modules and returns the aggregated results to the GPU Node over RDMA.

\subsection{PNM Software Stack}
The PNM software stack is illustrated in Figure~\ref{fig_pnm_dev}, organized into three layers: the NelssaHost, the PNM User Library, and the OS/HW layer. 
During prefill, the NelssaHost's KV Partitioner distributes the clustered KV cache across $N$ PNM devices via RDMA. During decode, the Job Dispatcher concurrently issues compute commands to all PNM devices, and the Output Aggregator collects the resulting partial output vectors and LSE values, merges them via online softmax, and returns the final output to the Decode GPU Node. Since only output vectors and LSE values are aggregated, merging overhead remains small and independent of KV cache size.

The PNM User Library exposes a Memory API for KV cache allocation and a Command API for compute operations. 
To balance prefill data ingestion and decode computing, \nelssa relies on API-level structural isolation rather than complex dynamic scheduling.
The Memory API handles prefill data distribution by bypassing the device's compute pipeline entirely. Meanwhile, the Command API independently manages decode execution, ensuring that the dedicated CMD Executor threads remain uninterrupted by background memory tasks.
On the device side, the Device Runtime interprets command descriptors and assigns a dedicated CMD Executor thread per descriptor for parallel execution, with each executor running the target kernel and returning the result upon completion. 

\section{Evaluation Methodology}\label{eval_method}
\niparagraph{Hardware configuration.} The Decode GPU Node consists of an Intel Xeon Platinum 8452Y CPU, 1 TB DDR5-4800 memory, and one or two NVIDIA H100 NVL GPUs (94 GB each). In the 2× GPU configuration, the two GPUs are connected via NVLink. The PNM Node, connected to the Decode GPU Node via a commodity Ethernet fabric, pairs an Intel Xeon 6952P CPU with four CXL-attached PNM devices, each equipped with four channels of on-board DDR5-6400 DRAM. 
In the current prototype, each PNM device is limited to 32 GB of device memory due to the on-board DRAM capacity. We accordingly report projected throughput at a 512 GB per-device configuration, consistent with commodity DDR5 platforms and the full-capacity PNM board we are currently developing. 
Table \ref{tab:my-table} details both hardware configurations. The projected model assumes only a memory expansion to 512 GB, keeping all other specifications identical to the prototype.
The projection methodology and its validation are described in Section~\ref{sec7.2}.



\begin{table}[t]
\centering
\resizebox{\columnwidth}{!}{%
\renewcommand{\arraystretch}{1.15}
\begin{tabular}{c|c|c}
\hline
\textbf{Component} &
\textbf{Prototype (Measured)} &
\textbf{Projected} \\
\hline
Compute Core       & 16x ARM Neoverse V2 & 16x ARM Neoverse V2 \\
Interconnect       & CXL 2.0 x16         & CXL 2.0 x16 \\
Memory Interface   & 4-CH DDR5-6400      & 4-CH DDR5-6400 \\
Memory Bandwidth   & 200 GB/s            & 200 GB/s \\
Memory Capacity    & 32 GB               & 512 GB \\
\hline
\end{tabular}
}
\caption{Hardware specifications for the measured prototype and projected configuration.}
\vspace{-10mm}
\label{tab:my-table}
\end{table}

\niparagraph{Workloads.} We evaluate on Llama3-8B-1048K \cite{llama-1048} across sequence lengths from 128K to 1024K tokens. Accuracy is measured using the RULER benchmark \cite{ruler}. For mixed-length serving evaluation, we construct a synthetic workload by sampling requests from the Mooncake conversation trace \cite{mooncake} under a Poisson arrival process, configured to reflect target mixed-length scenarios.

\niparagraph{Baselines.} We compare against three baselines. GPU-only (1× H100 NVL) and GPU-only (2× H100 NVL, NVLink) both execute dense decode attention via FlashAttention without any offloading. The 2×GPU baseline is included as a cost-comparable reference to account for the additional hardware resources in the \nelssa configuration. RetroInfer \cite{retroinfer} is a state-of-the-art CPU offloading system with sparse attention. \nelssa is evaluated as an end-to-end system comprising the Decode GPU Node and the PNM Node.

\niparagraph{Metrics.} The primary metric is decode throughput in tokens per second. We additionally report tail latency (P95 and P99 TPOT) and GPU memory utilization as secondary metrics.

\section{Experimental Results}
\subsection{End-to-End Throughput Evaluation}\label{sec7.2}
\niparagraph{Decode throughput.} 
Figure~\ref{fig_throughput3} compares decode throughput across baselines and \nelssa over varying sequence lengths and batch sizes. All experiments are conducted at a 4\% selection ratio, which ensures stable accuracy as established in Section~\ref{sec7.3}. Within the directly measured range (up to batch 8 at 128K, batch 4 at 256K, batch 2 at 512K, and batch 1 at 1M tokens), \nelssa serves all configurations where GPU-only baselines fail due to memory exhaustion, while sustaining throughput comparable to the GPU-only baselines even on our capacity-constrained prototype. Beyond this range, we project throughput to the full-capacity 512 GB per-device configuration described in Section~\ref{eval_method}. The throughput advantage grows with sequence length and batch size, as larger workloads improve PNM compute efficiency and reduce the relative impact of fixed per-step costs. Across the full evaluation range, \nelssa achieves up to 5.5$\times$ higher throughput than the 1× GPU baseline and up to 3.6$\times$ higher throughput than the 2× GPU baseline.

\begin{figure}[!t]
\centering
\includegraphics[width=1.0\columnwidth]{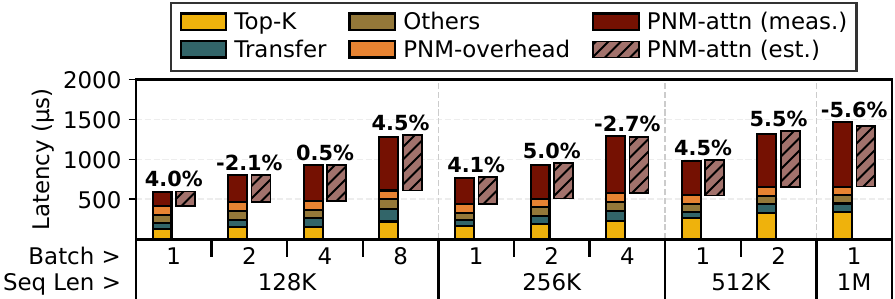}
\caption{Component-level breakdown of decode attention latency per step across sequence lengths and batch sizes. Discrepancy between measured (meas.) and estimated (est.) values for PNM-attn is within ±5.6\%.}
\label{fig_throughput2}
\end{figure}

\begin{figure}[!t]
\centering
\includegraphics[width=1.0\columnwidth]{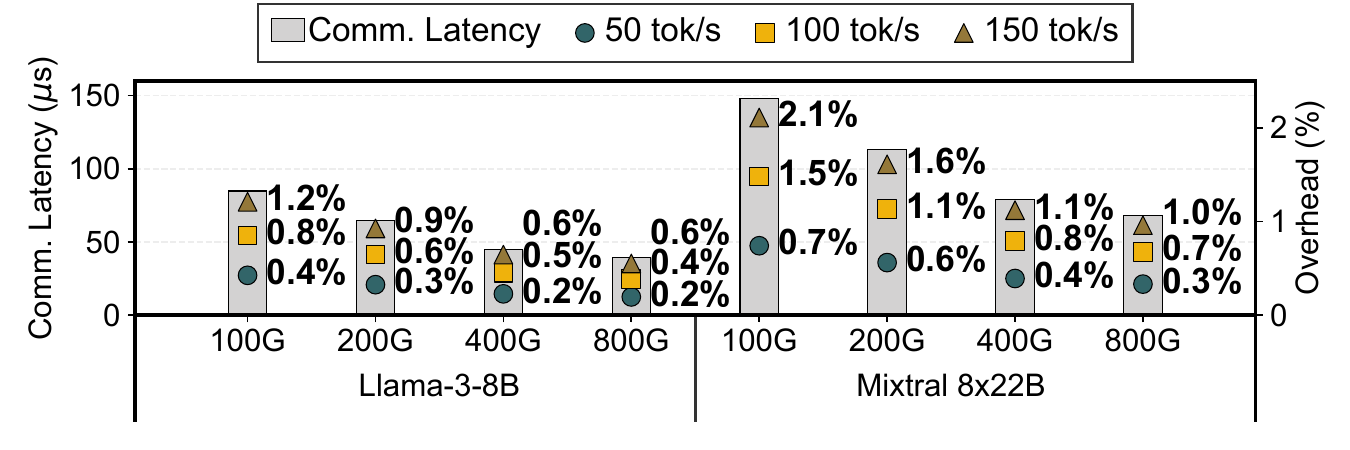}
\caption{GPU-PNM cross-node communication overhead as a fraction of total decode latency across different target throughputs and network configuration (analytical model).}
\label{fig_rev_comm}
\end{figure}

\niparagraph{Latency breakdown and projection.}
Figure~\ref{fig_throughput2} presents a component-level breakdown of the measured decode attention latency per step. The five components are Top-K similarity search, Transfer, PNM-overhead, PNM-attn, and Others, all of which are directly measured on the hardware prototype. For PNM-attn, estimated values derived from effective bandwidth and data volume are shown alongside measured values, with a discrepancy within ±5.6\%. The projection is constructed from this breakdown. Top-K values are taken from direct measurements across the full projected range. Transfer and Others are treated as constants using their measured averages within the evaluated range. PNM-overhead may diminish as data volume grows, but its measured average is held fixed as a conservative estimate. For PNM-attn, measurements show a tendency for effective bandwidth to increase with data volume, but the effective bandwidth observed within the measured range is held constant across the full projected range.
As only PNM-attn relies on estimation, the projection error of the overall breakdown is dominated solely by this term.

\niparagraph{Network sensitivity.}
We analyze the GPU-PNM cross-node communication overhead under 100G–800G network configurations (Figure \ref{fig_rev_comm}). Even under a stringent 150 tok/s target, which is a challenging condition for long-context workloads, the communication accounts for at most 2.1\% of the total decode latency. Due to the minimal per-step data exchange, \nelssa structurally avoids network bottlenecks despite requiring per-step coordination.

\begin{figure}[!t]
\centering
\includegraphics[width=0.95\columnwidth]{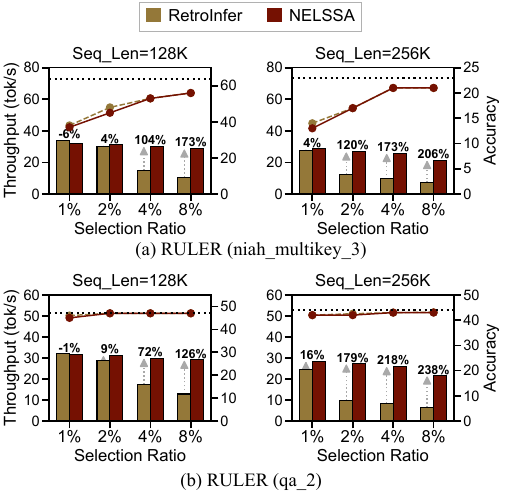}
\caption{Decode throughput and accuracy under varying selection ratios (Llama3-8B-1048K).}
\label{fig_main_sel_ratio}
\end{figure}

\subsection{Accuracy and Throughput under Varying Sparsity}\label{sec7.3}
\textbf{Accuracy.} Figure~\ref{fig_main_sel_ratio} presents decode throughput and accuracy across selection ratios on RULER benchmark tasks. At a selection ratio of 4\% and above, \nelssa maintains stable accuracy across both tasks while sustaining competitive throughput, confirming that this range provides a stable accuracy-throughput tradeoff. The decode throughput evaluation in Section~\ref{sec7.2} is accordingly conducted at a 4\% selection ratio.

\niparagraph{Throughput under varying sparsity.}
\nelssa exhibits a gradual and stable throughput trend as the selection ratio increases. Because attention computation completes entirely within PNM and only the result is returned, changes in the selection ratio do not directly affect interconnect traffic. In contrast, CPU offloading approaches transfer a proportionally larger volume of KV vectors across PCIe as the selection ratio increases, exposing them to interconnect pressure as a structural constraint. Throughput in such approaches can also vary depending on workload characteristics. As a result, CPU offloading approaches tend to achieve stable throughput only at low selection ratios, whereas \nelssa sustains a broad region where both accuracy and throughput requirements are jointly satisfied across a wide range of selection ratios.

\subsection{Head-of-Line Blocking Mitigation}\label{hol_sec}
\niparagraph{Evaluation methodology.} 
To evaluate the system-wide impact of \nelssa's hybrid routing, we employ a two-stage methodology combining direct hardware measurement with measurement-grounded system-level emulation. (i) GPU memory utilization, running batch size, and tail latency (P95/P99 TPOT) are measured directly on our hardware prototype, providing empirical evidence that \nelssa preserves GPU memory headroom and sustains higher running batch sizes under mixed-length workloads. (ii) For end-to-end decode throughput, we adopt an emulation methodology grounded in direct hardware measurements. The key insight is that the impact of \nelssa on GPU-side throughput is governed by its GPU memory and compute occupancy patterns, both of which are directly measurable on our prototype. We construct a workload injection sequence calibrated to these measured values and inject it into an unmodified vLLM engine, preserving its original scheduling, batching, and memory management behaviors so that request-level interactions are faithfully reproduced. Since the injected workload is constructed from requests at approximately 100K tokens, validation is performed at this length against direct hardware execution, confirming a measurement error of 4.12\%. The directly measured micro-metrics independently corroborate the same architectural trends, making the two methods mutually reinforcing in characterizing the system-level benefits of our approach.

\begin{figure}[!t]
\centering
\includegraphics[width=1.0\columnwidth]{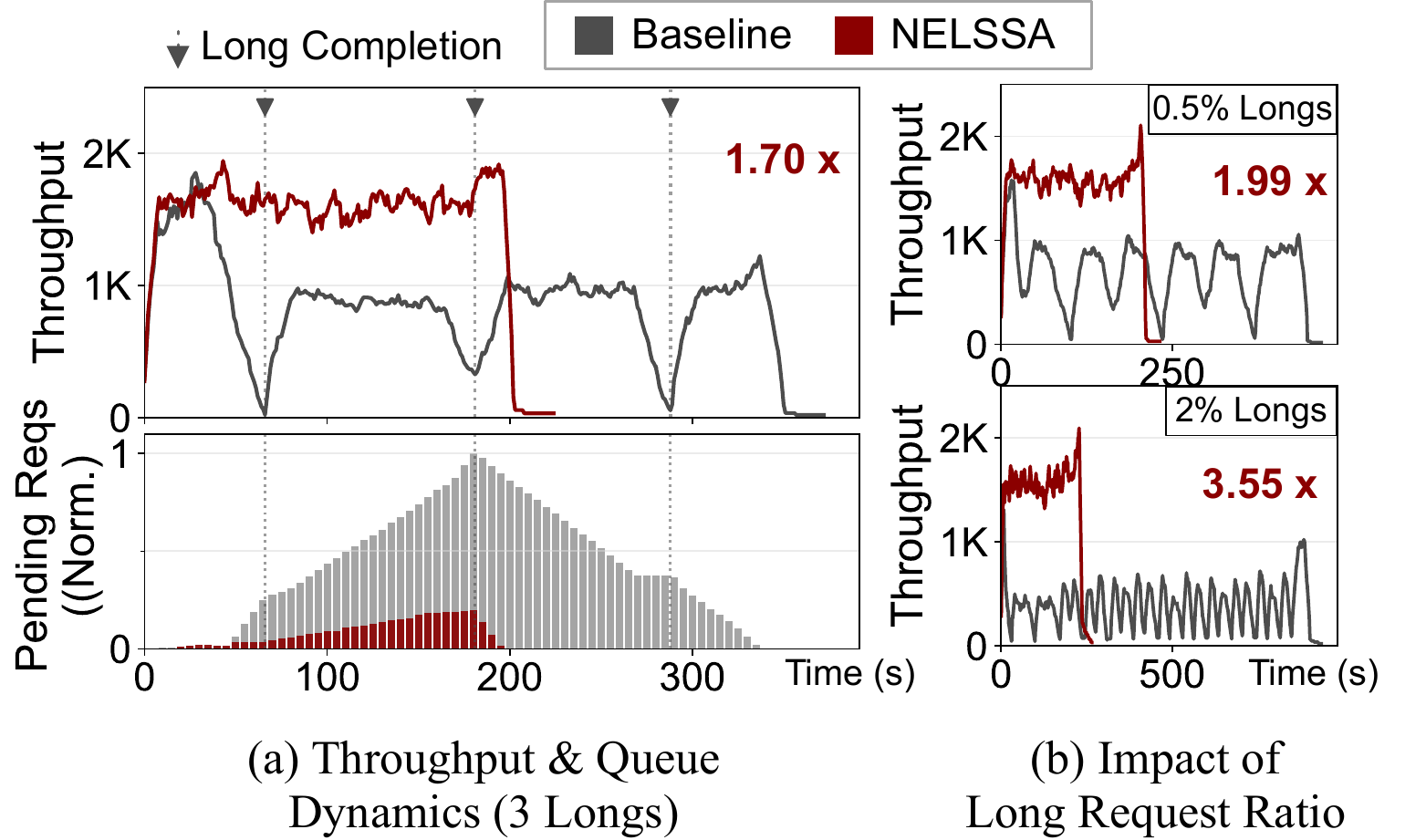}
\caption{System-wide decode throughput and pending queue dynamics (LLaMA-3-8B). (a) \nelssa sustains throughput under long request injection while the baseline stalls. (b) \nelssa's throughput advantage grows as the fraction of long requests increases.}
\label{fig_hol_results}
\end{figure}

\begin{figure}[!t]
\centering
\includegraphics[width=1.0\columnwidth]{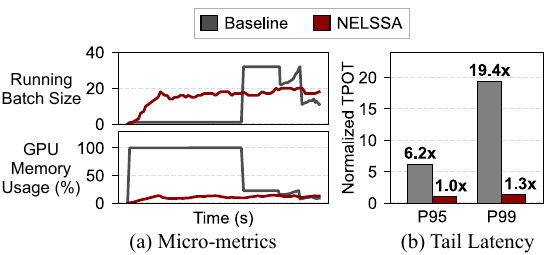}
\caption{Directly measured GPU memory utilization, running batch size, and tail latency upon injecting a single 256K long request (normalized to \nelssa P95). \nelssa preserves memory headroom and batch size, achieving 6.2$\times$ and up to 15$\times$ reduction in TPOT at P95 and P99, respectively.}
\label{fig_hol_results2}
\end{figure}

\begin{figure}[!t]
\centering
\includegraphics[width=1.0\columnwidth]{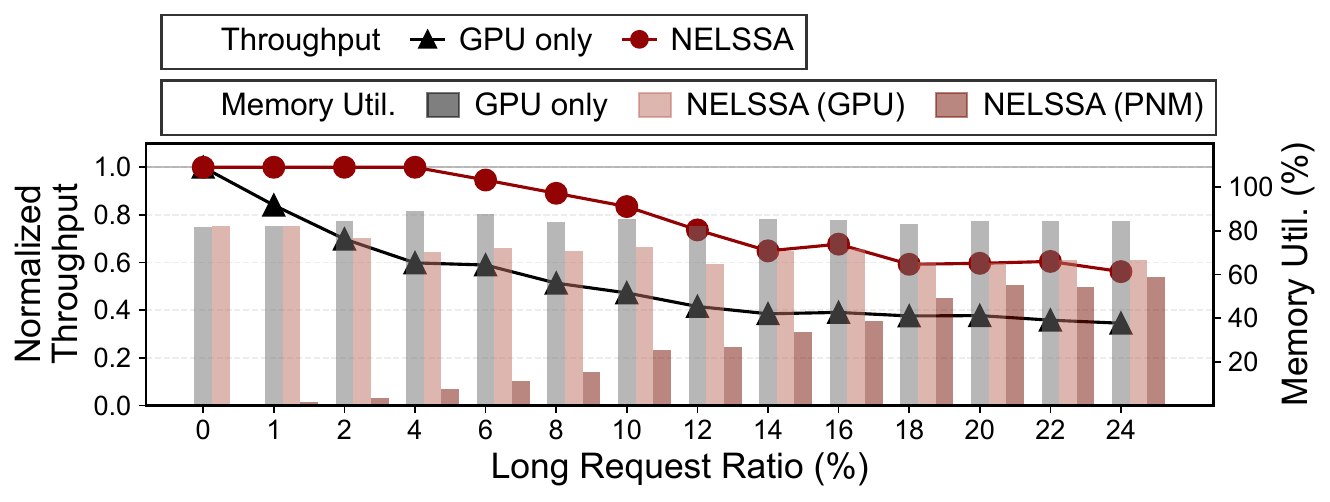}
\caption{Decode throughput under varying long request fractions from 0\% to 24\% for the baseline and \nelssa.}
\label{fig_rev_long_ratio}
\end{figure}

\niparagraph{Micro-metrics and tail latency.} The fundamental mechanism driving HoL blocking mitigation is evident in the directly measured micro-metrics (Figure~\ref{fig_hol_results2}). Upon the arrival of long requests, the baseline's GPU memory hits its capacity, forcing the scheduler to reduce the running batch size to as low as 1. In contrast, by decoupling the massive attention memory footprint, \nelssa maintains memory headroom on the GPU, seamlessly sustaining a higher running batch size. This architectural isolation directly translates into substantial improvements in the directly measured tail latency of short requests. Based on normalized Time-Per-Output-Token (TPOT) measurements, \nelssa achieves a 6$\times$ improvement at the 95th percentile (P95) and up to a 15$\times$ improvement at the 99th percentile (P99) compared to the baseline.

\niparagraph{Throughput and queue dynamics.} These micro-architectural differences directly manifest in the macroscopic throughput and pending queue dynamics, as illustrated in Figure~\ref{fig_hol_results}(a). Under mixed traffic (RPS=10), the baseline throughput experiences a notable drop upon long request injection, where the extent and duration of the throughput degradation vary depending on the burst size (e.g., BS=2 vs. BS=1). Furthermore, even after the long requests complete, the baseline fails to immediately restore its peak decode throughput. This sluggish recovery is primarily attributed to the overhead of swapping in previously evicted KV caches from the CPU and the scheduler's gradual restoration of the running batch size. Consequently, pending short requests rapidly accumulate to saturation (1.0). Conversely, \nelssa sustains consistently high throughput and averts queue accumulation.

Figure~\ref{fig_hol_results}(b) demonstrates \nelssa's robustness across diverse workload configurations with varying RPS and long request counts. While the baseline throughput is volatile under stress, \nelssa remains stable across the tested configurations. As the frequency of long requests increases (e.g., from 1 to 7 bursts, or reaching 2\% of total requests), the stalling in the baseline worsens, and \nelssa's relative throughput speedup shows an increasing trend. Considering the trajectory of future workloads--where the proportion and length of long-context requests are projected to increase--\nelssa's ability to fundamentally mitigate HoL blocking will be an increasingly important property for scalable LLM serving.

\niparagraph{Long request sensitivity.} To stress \nelssa under skewed request mixes, we fix the arrival rate and sweep the fraction of long-context requests from 0\% to 24\%. Figure~\ref{fig_rev_long_ratio} shows that the GPU-only baseline quickly loses throughput as long requests consume HBM and collapse the running batch size. In contrast, \nelssa’s throughput degradation occurs only at a much higher long-request fraction and remains well above the baseline throughout the evaluated range, demonstrating robustness to skewed workload mixes.

\begin{figure}[!t]
\centering
\includegraphics[width=1.0\columnwidth]{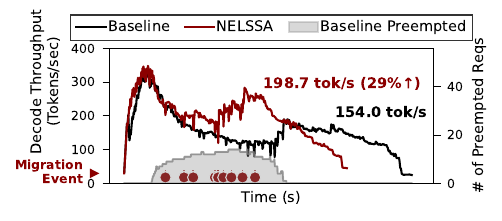}
\caption{System-wide decode throughput and preemption count under dynamic context growth. \nelssa sustains stable throughput via seamless background migration while the baseline suffers repeated preemptions.}
\label{fig_dynamic1}
\end{figure}

\subsection{Seamless Migration under Dynamic Context Growth}
\niparagraph{Evaluation methodology.} 
To evaluate \nelssa's effectiveness under dynamic context growth, we configure a synthetic workload with an input-to-output length ratio of 1:2, reflecting long-output generation patterns common in workloads such as reasoning and chain-of-thought inference.
In this setting, the number of output tokens is difficult to predict in advance \cite{response, output1, output2, output3, output4, output5, output6, output7}, causing frequent GPU memory saturation and out-of-memory events in the baseline. When GPU memory utilization reaches the migration threshold, \nelssa applies its migration policy to transfer the KV cache of requests exceeding the threshold to the PNM node via seamless background migration, allowing inference to continue without interruption. System-wide throughput trends are captured using the same emulation methodology as Section~\ref{hol_sec}, while migration overhead is directly measured on the hardware prototype.

\begin{figure}[!t]
\centering
\includegraphics[width=1.0\columnwidth]{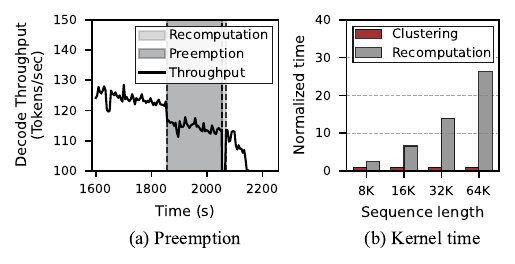}
\caption{(a) Throughput degradation at preemption events in the baseline. (b) Recomputation latency vs. incremental clustering overhead, directly measured on hardware.}
\label{fig_dynamic2}
\end{figure}

\begin{figure}[!t]
\centering
\includegraphics[width=\columnwidth]{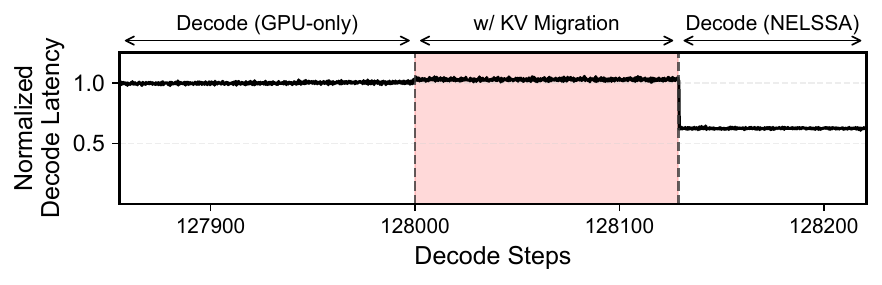}
\caption{Decode latency jitter during concurrent KV cache migration (microbenchmark at 128K sequence length).}
\label{fig_rev_jitter}
\end{figure}

\niparagraph{Throughput and migration overhead.}
Figure~\ref{fig_dynamic1} shows system-wide throughput under the dynamic growth scenario. In the baseline, GPU memory saturation causes the running batch size to shrink progressively, increasing the number of preempted requests and reducing overall throughput. 
\nelssa reclaims GPU memory headroom through seamless background migration and sustains a stable running batch size throughout the entire execution. In this scenario, token generation time accounts for the majority of end-to-end latency, which naturally limits the relative throughput gain from migration. Nevertheless, \nelssa achieves a 29\% improvement over the baseline.

Furthermore, a 128K sequence microbenchmark (Figure~\ref{fig_rev_jitter}) shows that background KV cache transfers cause no significant decode latency jitter. This confirms concurrent migration introduces minimal interference to active inference.

Figure~\ref{fig_dynamic2} examines the underlying cause of this throughput difference. Figure~\ref{fig_dynamic2}(a) shows that preemption events in the baseline cause noticeable throughput degradation. As shown in Figure~\ref{fig_dynamic2}(b), this stems from recomputation latency, which grows steeply with context length. In contrast, \nelssa's incremental clustering overhead remains relatively insensitive to context length, a property confirmed through direct hardware measurement. Taken together, these results demonstrate that the one-way handoff principle in \nelssa extends effectively to dynamically growing requests, sustaining stable execution even when context length exceeds GPU memory capacity at runtime.



\vspace{4mm}
\section{Related Work}
\subsection{Algorithmic KV Cache Management}
Algorithmic approaches to managing long-context KV caches fall into two categories. KV compaction and quantization methods \cite{streamingllm, H2O, kvzip, duoattention, infiniplot, fastkv} reduce HBM footprint by pruning or summarizing less important tokens, but discard information in a query-agnostic manner, leading to accuracy degradation that worsens as the compression ratio increases. Sparse attention approaches with CPU offloading \cite{flexgen, infinigen, retrievalattention, lmcache, arkvale, retroinfer} avoid this loss by retaining the full KV cache in host memory, but face an unavoidable tradeoff: increasing the selection ratio raises PCIe bandwidth pressure, while reducing it degrades accuracy. In both categories, a single parameter (compression ratio or selection ratio) jointly governs accuracy and throughput, and this tradeoff becomes more sensitive under varying workload characteristics.


\subsection{Sparse Attention on PNM}
Early PNM-based efforts \cite{cxl_pnm, hermes} demonstrated the viability of near-memory attention execution but remained limited in throughput by lower memory bandwidth than GPU HBM. To address this, subsequent work proposed combining sparse attention with PNM execution \cite{cal_nelssa, longsight, scalable_pnm}, and the prior work \cite{cal_nelssa} demonstrated through simulation that sparsity can effectively mitigate the capacity-bandwidth tradeoff.

LongSight \cite{longsight} and ScalablePNM \cite{scalable_pnm} propose more concrete architectures in this direction, but differ fundamentally from \nelssa in execution model. LongSight \cite{longsight} performs filtering and Top-K selection within a compute-enabled memory device, but transfers the selected Top-K QKT and Value vectors to the GPU for final attention computation, leaving interconnect overhead unresolved and constraining the selection ratio through hardware limits. ScalablePNM \cite{scalable_pnm} executes the full pipeline from token selection to attention computation entirely within PNM, but relies on a custom LPDDR5X-based PNM platform delivering 1.1 TB/s per module and is validated only through architectural simulation. On commodity DRAM, the bandwidth available for near-memory processing is substantially lower, and performing token selection within PNM under this constraint introduces non-trivial decode latency overhead in practical deployments. \nelssa accounts for this by assigning token selection to GPU-side centroid search and confining attention execution to PNM, arriving at a practical operating point that satisfies both latency and accuracy requirements on commodity DRAM. This placement boundary is derived from hardware performance characteristics rather than set heuristically.

Both LongSight \cite{longsight} and ScalablePNM \cite{scalable_pnm} assume static request lengths and do not address system-level challenges such as mixed-length workload heterogeneity or dynamic context growth in real serving environments. To the authors' knowledge, \nelssa is the first end-to-end system implemented on real PNM hardware that jointly addresses these challenges within a unified serving stack.

\section{Discussion}
While \nelssa's length-aware routing provides a stable first-order baseline, unpredictable workload skews in real-world multi-tenant environments may saturate specific hardware tiers.
To address this, we propose extending the system with orchestrator-level adaptive routing as future work. Positioned above the large-scale LLM serving pipeline, a cluster orchestrator can monitor queueing delays and node utilization in real time. It functions as a traffic valve, dynamically adjusting the routing threshold ($T_{input}$) when resource saturation is anticipated.
Because \nelssa is designed as a modular add-on to existing prefill-decode disaggregated infrastructures, these software-level dynamic scheduling techniques can be seamlessly integrated into production environments without disrupting existing data center topologies.

\section{Conclusion}
This work builds \nelssa, a GPU–PNM heterogeneous LLM serving system for mixed-length workloads, introducing length-based request placement with a hardware-aware crossover threshold and a seamless one-way migration mechanism that enables execution handoff under dynamic context growth without recomputation.
\nelssa underscores a broader architectural shift: as agentic AI drives increasingly mixed and long-context workloads, KV cache capacity and memory efficiency are emerging as primary scalability bottlenecks in LLM serving. 
Simply scaling GPUs or relying on compression techniques is unlikely to provide a sustainable path forward. 
Our results suggest that heterogeneous execution tiers, where memory-centric accelerators such as PNM serve as first-class decode engines rather than auxiliary offload devices, offer a practical and scalable direction for future LLM infrastructure.
%


\bibliographystyle{ACM-Reference-Format}
\bibliography{reference}

\end{document}